\documentstyle[prd,aps]{revtex}
\def\be{\begin{equation}}
\def\ee{\end{equation}}
\def\ba{\begin{eqnarray}}
\def\ea{\end{eqnarray}}

\begin{document}
%
%
\draft
\title{Matter effects in the $D^0 - \overline{D^0}$ system}
\author{Jo\~ao P.\ Silva\footnote{Permanent address: 
	Instituto Superior de Engenharia de Lisboa,
	Rua Conselheiro Em\'{\i}dio Navarro,
	1900 Lisboa, Portugal.}}
\address{Stanford Linear Accelerator Center, Stanford University, Stanford,
	 California 94309}
\date{\today}
\maketitle
\begin{abstract}
We discuss the impact of matter effects in the $D^0 - \overline{D^0}$ system.
We show that such effects could, in principle, be measured,
but that they cannot be used to probe the mass difference $x_D$ or
the lifetime difference $y_D$.
This occurs because the mixing effects and the matter effects decouple
at short times.
We also comment briefly on the $B$ systems.
\end{abstract}


\section{Introduction}

The $D^0 - \overline{D^0}$ system is of great current interest.
Until recently,
the best probes on the mixing resulted only in upper bounds on 
$x_D^2 + y_D^2$ of order $10^{-2}$ \cite{PDG}.
This changed with the recent results from CLEO \cite{CLEO} and
FOCUS \cite{FOCUS}.
The current situation is beautifully summarized in reference
\cite{Ber00}:
there is evidence for $y_D \sim 10^{-2}$ and for a large strong phase,
corresponding to large SU(3) breaking effects.
The problem with the strong, final-state interaction phases is that
they are not known.
They are fit in the same experiments that look for $y_D$.
It would be interesting if matter effects could be used to shed some
light on this issue.
The idea arises from the kaon system,
where the phases due to matter effects are measured experimentally.
Thus,
we would be dealing with a {\em known} CP-even phase.

There is one important difference between the kaon and $D$ systems.
For a beam of neutral kaons in vacuum,
the depletion from the beam is controlled by two exponentials,
$\exp{(- \Gamma_S\, t)}$ and $\exp{(- \Gamma_L\, t)}$,
corresponding to the short-lived ($K_S$) and long-lived ($K_L$)
components.
Here $|\Delta \Gamma| \sim 2 \Gamma$ and the
two exponential fallouts are clearly separated.
As a result,
we may wait for the $K_S$ component to decay away and use the
matter effects on $K_L$ to regenerate $K_S$;
a phenomenon known as `regeneration'.
In the $D$ (and the $B_d$) system the situation is very different:
$|\Delta \Gamma| \ll \Gamma$ and the 
leading behavior in vacuum is given by $\exp{(- \Gamma\, t)}$,
with the hyperbolic sine and cosine of $\Delta \Gamma\, t$
acting as small perturbations.
In this system,
the depletion from the beam is controlled by $\Gamma$.
Because we cannot resolve the two exponentials,
the classical `regeneration' experiments cannot be carried out
and a new analysis is required.
This is the question we address here.

In section~\ref{sec:propagation} we review the general features
of propagation of neutral mesons systems in matter.
In section~\ref{sec:DDbar} we use an extrapolation from known
kaon results to argue that these effects might,
in principle,
be sought.
Unfortunately,
we will also show that,
even if visible,
the matter effects cannot be used to probe $x_D$ or $y_D$.
This is shown to be a generic feature of the `small-time' approximation
in the time-evolution.
We draw our conclusions in section~\ref{sec:conclusions},
where we remark briefly on the $B_d$ and $B_s$ systems.

\section{Propagation in matter}
\label{sec:propagation}

Let us consider a generic neutral meson system
$P^0 - \overline{P^0}$.
Assuming CPT-conservation in vacuum,
this system is characterized by two complex eigenvalues,
$\mu_a = m_a - i\Gamma_a/2$ and $\mu_b = m_b - i\Gamma_b/2$,
and by a mixing parameter,
$q/p$.
It is convenient to define the average and the difference of the
eigenvalues,
\ba
\mu = \frac{\mu_a + \mu_b}{2}
&=&
m - \frac{i}{2} \Gamma,
\nonumber\\
\Delta \mu = \mu_a - \mu_b
&=&
\Delta m - \frac{i}{2} \Delta \Gamma.
\ea
These quantities describe the system in vacuum and its time evolution
is well known \cite{BLS}.

The new effects resulting from the interaction with matter can be
described by the elastic forward scattering amplitudes of $P^0$
and $\overline{P^0}$,
which we denote by $f$ and $\bar f$,
respectively.
These enter the effective Hamiltonian through
\be
\chi = - \frac{2 \pi N}{m} f,
\ \ \ \mbox{and}\ \ \ 
\bar \chi = - \frac{2 \pi N}{m} \bar f,
\ee
where $N$ is the density of scattering centers in the medium.
As before,
we define the average and the difference of these parameters
as
\ba
\chi_{\rm av} &=& \frac{\chi + \bar \chi}{2},
\nonumber\\
\Delta \chi &=& \chi - \bar \chi = - \frac{2 \pi N}{m} \Delta f,
\label{definition-chi}
\ea
with $\Delta f = f - \bar f$.

In the presence of matter,
the time evolution of meson states with a given initial flavor may
be written as \cite{early,Fet96,Sch98,joao},
\ba
e^{i(\mu+\chi_{\rm av})t}\,
|P^0(t)\rangle & = &
\left[ \cos{\left(\Omega\, t/ 2 \right)}
- i \frac{\Delta \chi}{\Omega}
\sin{\left(\Omega\, t/2 \right)}
\right]
|P^0 \rangle
+
\frac{q}{p}
\left[
-i  \frac{\Delta \mu}{\Omega}
\sin{\left(\Omega\, t/2 \right)}
\right]
|\overline{P^0} \rangle,
\nonumber\\*[3mm]
e^{i(\mu+\chi_{\rm av})t}\,
|\overline{P^0}(t)\rangle & = &
 \frac{p}{q} 
\left[
-i \frac{\Delta \mu}{\Omega}
\sin{\left(\Omega\, t/2 \right)}
\right]
|P^0 \rangle
+
\left[ \cos{\left(\Omega\, t/2 \right)}
+ i \frac{\Delta \chi}{\Omega}
\sin{\left(\Omega\, t/2 \right)}
\right]
|\overline{P^0} \rangle,
\label{master}
\ea
where $\Omega = \sqrt{(\Delta \mu)^2 + (\Delta \chi)^2}$ \cite{definition}.
The evolution in vacuum is reproduced by setting
$\chi_{\rm av}$ and $\Delta \chi$ to zero.

As mentioned in the introduction,
in the $D$ (and in the $B_d$) system the depletion from the beam in
vacuum is controlled by $\Gamma$.
This {\em one} exponential is then modulated by
trigonometric functions of $\Delta \mu\, t = \frac{\Delta \mu}{\Gamma}\, \tau$,
where $\tau = \Gamma\, t$ is the time measured in lifetime units.
Therefore,
one introduces
\be
\frac{\Delta \mu}{\Gamma} = x - i y.
\label{xandy}
\ee
In most cases of interest the matter effects satisfy
$\Gamma > |\mbox{Im} \chi_{\rm av}|$ and
$|\Delta \mu| > |\Delta \chi|$,
and the depletion is still chiefly determined by $\Gamma$.
To describe the trigonometric functions in matter
we need Eq.~(\ref{xandy}) and a new complex parameter.
This may be choosen to be $\Delta \chi/\Gamma$,
which determines the depletion (statistics available)
by the time the effects of $\Delta \chi$ become important.
However,
since we know the current experimental reach on $\Delta \mu$,
it is easiest to use it to access the experimental reach
on $\Delta \chi$,
through\footnote{Although it is conventional to use this factor of
two in the definition of $r$,
it seems to confuse rather than enlighten the issue.}
\be
2 r = \frac{\Delta \chi}{\Delta \mu}.
\ee
Our figure of merit will be $|\Delta \chi/\Delta \mu|$.

Henceforth we will refer to those effects due to $\Delta \mu$
as `mixing effects',
and to those effects due to $\Delta \chi$
as `matter effects'.
The fundamental question is:
when do the matter effects and the mixing effects couple to each other?
The answer is easily found by looking at Eqs.~(\ref{master}).
They can couple in two ways.
The first and simplest way occurs when the production, propagation and
decay occur all in the same medium and the decay is flavor tagging.
In that case,
the matter and mixing effects couple mainly through $\Omega$.
The second way occurs when both flavors can decay into the
same final state,
or when the system crosses several media.
For example,
for events occuring completely in the medium,
with the decays going into CP-eigenstates.
Or,
when the production and the initial part of the evolution
occur in matter,
with the final part of the evolution and its decay occuring
in vacuum.
In these cases,
besides $\Omega$,
there can also be effects proportional to $\Delta \mu\, \Delta \chi$,
which arise from the interference between the coefficients
of the sine terms multiplying $|P^0 \rangle$ and of 
the sine terms multiplying $|\overline{P^0} \rangle$.

\section{Matter effects in the $D^0 - \overline{D^0}$ system}
\label{sec:DDbar}

\subsection{Interplay between mixing and matter effects}

We will now show that,
even when visible,
the matter effects cannot be used to probe $x_D$ or $y_D$.
The main reason is that the complex trigonometric functions
are controled by $(x - i y) \tau$ and,
with current technology,
one can only follow $\tau$ up to 5 or 10,
corresponding to $5$ to $10$ average lifetimes in the 
$D$ proper frame.
This is mostly due to the statistics of perfectly reconstructed events.
It is obvious that,
given unlimmited event samples,
one could follow the full decay curve including times of order $1/y_D$.
We mention this because the `no-go theorem' we are about to present
hinges on the experimental constraint that one can only
follow the decay curves for time $\tau \ll 1/y_D$.
Under these circumstances,
we can expand the trigonometric functions on the right-hand side
of Eqs.~(\ref{master}) as
\be
\cos{\left(\Omega\, t/ 2 \right)}
\pm i \frac{\Delta \chi}{\Omega}
\sin{\left(\Omega\, t/2 \right)}
\approx
1
\pm i \frac{\Delta \chi}{2}\, t
- \frac{(\Delta \mu)^2 +(\Delta \chi)^2 }{8}\, t^2
\mp i \Delta \chi\, \frac{(\Delta \mu)^2 +(\Delta \chi)^2}{48}\, t^3
+
\bigcirc (t^4),
\label{ye-1}
\ee
and
\be
-i \frac{\Delta \mu}{\Omega}
\sin{\left(\Omega\, t/2 \right)}
\approx
- i \frac{\Delta \mu}{2}\, t
+ i \Delta \mu\, \frac{(\Delta \mu)^2 +(\Delta \chi)^2}{48}\, t^3
+
\bigcirc (t^5),
\label{ye-2}
\ee
for the right-sign and wrong-sign transitions,
respectively.

The old results on semileptonic wrong-sign decays \cite{E791}
measure the magnitude squared of Eq.~(\ref{ye-2}),
given to leading order by
\be
\frac{|\Delta \mu|^2}{4 \Gamma^2}\, \tau^2 = \frac{x^2 + y^2}{4} \tau^2.
\label{quadratic-delta-mu}
\ee
Let us now consider a final state $f$ to which both $P^0$
and $\overline{P^0}$ can decay.
We denote the decay amplitudes by $A_f$ and $\bar A_f$,
respectively.
Such a decay has a term linear in $t$ arising from the interference
between Eqs.~(\ref{ye-1}) and (\ref{ye-2}),
\be
\mbox{Im}
\left[ \Delta \mu\, \frac{q}{p} \bar A_f A_f^\ast \right]\, t.
\label{linear-delta-mu}
\ee
CLEO \cite{CLEO} has looked at $f = K^+ \pi^-$ which is
a Cabibbo-allowed decay for $\overline{D^0}$ and a
Doubly-Cabibbo-suppressed decay for $D^0$.
FOCUS \cite{FOCUS} has looked at the CP-eigenstate $f=K^+ K^-$.

Could one look at matter effects with semileptonic decays?
In principle yes.
The time evolution of the right-sign semileptonic decays is proportional
to the magnitude squared of Eq.~(\ref{ye-1}),
which has a term linear in $t$,
\be
\mp \mbox{Im} \left( \frac{\Delta \chi}{\Gamma} \right)\, \tau.
\ee
Even if $|\Delta \chi|$ is an order of magnitude smaller
than $|\Delta \mu|$,
this term should still be easier to detect than
the quadratic term in Eq.~(\ref{quadratic-delta-mu}).
This idea will be explained in detail below.

For the moment we wish to return to our fundamental question:
can matter effects be used to give a new handle on $\Delta \mu$?
The answer is no! First consider right-sign semileptonic decays
and magnitude square the right-hand side of Eq.~(\ref{ye-1}).
We see that only at order $\tau^3$ do we find effects proportional
to $\Delta \mu\, \Delta \chi$.
Even at order $\tau^2$,
the effects decouple as $\mbox{Re} (\Delta \mu)^2$ and 
$\mbox{Re} (\Delta \chi)^2$.
Next consider wrong-sign semileptonic decays
and magnitude square the right-hand side of Eq.~(\ref{ye-2}).
This case is even worse,
since the effects proportional to $\Delta \mu\, \Delta \chi$
only arise at order $\tau^4$.
Now consider again a final state $f$ to which both $D^0$
and $\overline{D^0}$ can decay.
In the presence of matter Eq.~(\ref{linear-delta-mu})
gets changed into
\be
\mbox{Im}
\left[ \Delta \mu \, \frac{q}{p} \bar A_f A_f^\ast 
\left(
t
+
i \frac{\Delta \chi^\ast}{2} t^2
\right)
\right].
\ee
In this case there is an effect proportional to
$\Delta \mu \, \Delta \chi^\ast$ at order $t^2$.
But,
if $|\Delta \chi|$ is an order of magnitude smaller than $|\Delta \mu|$,
then this effect will be an order of magnitude smaller
than that in Eq.~(\ref{quadratic-delta-mu}).
Notice that,
in particular,
there is no effect proportional to $\Omega t$.

In this discussion we have implicitly considered only
events which occur totally in matter.
However,
the matter-vacuum transition can only promote the same type of interference
effects we have already discussed \cite{fine-tunning}.

\subsection{Observing matter effects}

We have already argued that matter effects cannot be used as a handle
on $x_D$ and $y_D$.
We have yet to prove that they are observable at all,
even in principle.
Let us consider the decay $D^0 \rightarrow K^- l^+ \nu_l$.
Using Eqs.~(\ref{master}) and (\ref{ye-1}) we find
\be
\Gamma \left[ D^0(t) \rightarrow K^- l^+ \nu_l \right]
\approx
e^{- \Gamma t}\, e^{2 \mbox{\small Im}\, \chi_{\rm av}\, t}\,
|A_o|^2
\left(
1
+
\mbox{Im}\, \Delta \chi\  t
\right),
\label{detect}
\ee
where we have denoted
$A_o = A(D^0 \rightarrow K^- l^+ \nu_l)$.\footnote{To illustrate this effect,
we have simplified the discussion by considering a `thought-experiment'
with the events taking place (production, evolution, and decay)
inside the material.
For these purposes,
a muon in the final state would be most readily observed.}
We now need to estimate the linear term.

However,
since the elastic forward scattering amplitudes for 
$D^0$ and $\bar{D}^0$ ($f$ and $\bar{f}$) are not known,
we have to estimate them somehow.
We recall that the matter effects in the kaon system arise because
$K^0$ interacts quasi-elastically with the nucleons,
while $\overline{K^0}$ suffers inelastic interactions,
such as
\ba
\overline{K^0}\ +\ (p^+, n)\ &\rightarrow& \pi^0 +\ (\Sigma^+, \Sigma^0)
\nonumber\\
s \bar d\ +\ (uud,udd) &\rightarrow& d \bar d\ +\ (uus,uds).
\label{inelastic-Kbar0}
\ea
The counterpart in the $D$ system is
\ba
D^0\ +\ (p^+, n)\ &\rightarrow& \pi^0 +\ (\Sigma_c^+, \Sigma_c^0)
\nonumber\\
c \bar u\ +\ (uud, udd)&\rightarrow& u \bar u\ +\ (udc, ddc).
\label{inelastic-D0}
\ea
The kinematics are different because
$m(D^0) \approx 1865\mbox{MeV} \gg m(K^0) \approx 498\mbox{MeV}$
and
$m(\Sigma_c^+) \approx 2454\mbox{MeV} \gg
m(\Sigma^+) \approx 1189\mbox{MeV}$.
However,
on the one hand we will be interested in the high momentum
range (larger than 25GeV),
where the difference becomes less important.
And,
on the other hand,
we are not interested in exact results but, rather,
wish to learn whether the matter effects are in principle observable.
Therefore,
we will simply scale the results from the kaon system to the
$D$ system.

We will use an empirical scaling law determined by Gsponer
and collaborators from their measurements in C, Al, Cu, Sn, and Pb for
kaon momenta between 30 to 150~GeV.
They found that \cite{Gsp79diff,Roe77}
\be
|\Delta f| = 1.13\mbox{fm} 
\left( \frac{A}{\mbox{g mol}^{-1}}\right)^{0.758} 
\left(\frac{p_K}{\mbox{GeV}c^{-1}}\right)^{0.386},
\label{Gsp-1}
\ee
where $A$ the atomic number of the material.
This result exhibits a power law momentum dependence
in accordance with Regge theory \cite{Regge},
which also predicts that $\arg{\Delta f}$ should be constant and
given by $-(1+0.386)\pi/2 = -0.693 \pi$ \cite{Roe77}.
The power-law approximation is fairly good down to a few GeV/$c$ momentum,
where low-energy resonances set in \cite{CPLEAR_disp}.
We neglect these effects since we will be concentrating on the
high momentum range.
We will also need the imaginary part of $f + \bar f$.
Again we rely on Gsponer and collaborators,
who found \cite{Gsp79sum,Thomas}
\be
\mbox{Im} (f + \bar f) = 
1.895\mbox{fm}\, 
\left( \frac{A}{\mbox{g mol}^{-1}}\right)^{0.840} 
\left(\frac{p_K}{\mbox{GeV}c^{-1}}\right)
\label{eq:thomas}
\ee
in the kaon sector.

We start by noticing that the 
density of scattering centers is a medium is given by
\be
N = \frac{N_{A} \rho}{A},
\ee
where $N_A$ is Avogadro's number,
$\rho$ is the density of the material,
and $A$ is its atomic number.
Using Eqs.~(\ref{definition-chi}) and (\ref{Gsp-1}),
we obtain
\ba
\frac{\Delta \chi_D}{\Gamma_D}
&=&
\frac{1}{m_D \Gamma_D} \left( - \frac{2 \pi N_A \rho}{A} \right)
\Delta f
\nonumber\\
&=&
-
5.6 \times 10^{-5} e^{i 0.307 \pi}
\left( \frac{A}{\mbox{g mol}^{-1}}\right)^{-0.242}
\left( \frac{\rho}{\mbox{g cm}^{-3}}\right)^{-0.242}
\left(\frac{p_D}{\mbox{GeV}c^{-1}}\right)^{0.386}.
\label{delta-chi-Gsp}
\ea
This result should be compared with that obtained for the neutral
kaons,
where $- 5.6 \times 10^{-5}$ is substituted by $0.09$.
The sign difference arises from the fact that 
the inelastic interactions occur for $D^0$,
not $\overline{D^0}$;
for example,
Eq.~(\ref{inelastic-D0})
involves $D^0$,
while Eq.~(\ref{inelastic-Kbar0}) involves $\overline{K^0}$.
Since we are assuming that the forward scattering amplitudes coincide,
the difference in magnitudes arises from
\be
\frac{|\Delta \chi_D|/\Gamma_D}{|\Delta \chi_K|/\Gamma_K}
=
\frac{m_K \Gamma_K}{m_D \Gamma_D}
=
\frac{498\mbox{MeV}}{1865\mbox{MeV}}
\frac{0.415 \times 10^{-12}s}{2 \times 0.893 \times 10^{-10}s}
= 6.2 \times 10^{-4}.
\label{scale}
\ee
The results for carbon and tungsten are listed in table~I
for a variety of $D$ momenta chosen within a range accessible to
FOCUS \cite{FOCUS-2}.
Tungsten is a preferable material since it leads to values
which are roughly $4.4$ times larger than those obtained in carbon.
For tungsten and momenta above $100 \mbox{GeV}c^{-1}$ the effects
are of order 0.2\%.
However,
the current experiments already probe $\Delta \mu_D/\Gamma_D$
of order 1\%.
Therefore,
in principle,
the linear term in Eq.~(\ref{detect}) would be detected with an experiment
of this type.

Of course,
we still need to show that the absorption term
$\exp{(2 \mbox{Im}\, \chi_{\rm av}\ t)}$ does not affect
greatly the number of events.
Using Eq.~(\ref{eq:thomas}),
we find
\be
2 \frac{\mbox{Im}\, \chi_{{\rm av}\ D}}{\Gamma_D}
=
- 9.4 \times 10^{-5}
\left( \frac{A}{\mbox{g mol}^{-1}}\right)^{-0.160}
\left( \frac{\rho}{\mbox{g cm}^{-3}}\right)
\left(\frac{p_D}{\mbox{GeV}c^{-1}}\right).
\label{chi-av-Gsp}
\ee
For tungsten and a momentum of $100 \mbox{GeV}c^{-1}$,
the ratio becomes $0.08$.
Looking back at the right-hand side of Eq.~(\ref{detect}),
this result means that the absorption exponential,
$\exp{(2 \mbox{Im}\, \chi_{\rm av}\, t)}$,
does not compete with the decay exponential,
$\exp{(- \Gamma t)}$.

Although we have been detailed about our estimates,
our objective is not to propose a specific experiment,
but rather to emphasize that one is not,
in principle,
out of reach.
Therefore,
we will not address the obvious list of questions:
i) How does one identify the primary and secondary vertices?;
ii) How does one instrument the material?;
iii) What is the ideal momentum range?;
etc.
We stress that the results presented here are based on the expectation
that we can extrapolate from the kaon to the $D$ system.
They should be taken as estimates and not as reliable predictions.
In particular,
the presence of resonances could enhance or suppress these estimates.

\section{Conclusions}
\label{sec:conclusions}

The situation in the $B_d$ system is similar to the one in the
$D$ system in that $\Delta \Gamma \ll \Gamma$.
Using the same naive high-momenta, scaling law used in Eq.~(\ref{scale}),
we obtain
\be
\frac{|\Delta \chi_B|/\Gamma_B}{|\Delta \chi_K|/\Gamma_K}
=
\frac{m_K \Gamma_K}{m_B \Gamma_B}
= 8.2 \times 10^{-4}.
\ee
However,
in the $B_d$ system $|\Delta \mu| \sim x_d \sim 0.7$,
meaning that $|\Delta \chi|$ cannot compete with $\Gamma$ nor with
$|\Delta \mu|$.

The situation in the $B_s$ is very different.
From the point of view of the lifetime difference it is 
actually closer in spirit to the kaon system \cite{Dun95}.
Indeed,
in the $B_s$ system $|\Delta \Gamma|/\Gamma \sim 0.15$
\cite{Ben96},
and one may talk about two exponentials.
However,
$B_s$ contains no quarks of the first family and, hence,
it would seem inappropriate to estimate the matter effects
as we have done above.

In conclusion,
we have analyzed the possibility of detecting matter effects in the
$D^0 - \overline{D^0}$ system.
We conclude that these effects are,
in principle,
accessible.
Unfortunately,
when the time dependence is followed only for a few lifetimes,
the matter effects and the mixing effects decouple.
As a result,
matter effects cannot be used to provide a new handle
on $x_D$ or $y_D$.

\acknowledgments

I am indebted to L.\ Lavoura and H.\ R.\ Quinn for reading the manuscript
and for useful suggestions.
I would like to thank H.\ R.\ Quinn,
Th.\ Schietinger,
and A.\ E.\ Snyder
for discussions and for their collaboration in a related subject.
This work is supported in part by the Department of Energy 
under contract DE-AC03-76SF00515.
The work of J.\ P.\ S.\ is supported in part by Fulbright,
Instituto Cam\~oes, and by the Portuguese FCT, under grant
PRAXIS XXI/BPD/20129/99	and contract CERN/S/FIS/1214/98.


\vspace{2cm}

%
\begin{table}[t]
\centering
\begin{tabular}{|c|c|c|}
\hfil $\frac{p_D}{\mbox{GeV}c^{-1}}$\hfil& 
\hfil $100
\left|\frac{\Delta \chi_D}{\Gamma_D}\right|_{\rm in\ carbon}$\hfil &
\hfil $100
\left|\frac{\Delta \chi_D}{\Gamma_D}\right|_{\rm in\ tungsten}$\hfil 
\\\hline
25& 
0.024&
0.106
\\
50& 
0.032&
0.138
\\
100& 
0.042&
0.182
\\
150& 
0.048&
0.212
\\
200& 
0.054&
0.236
\\
250& 
0.058&
0.258
\\
\end{tabular}
\caption{Values of $\Delta \chi_D/\Gamma_D$ for carbon and tungsten,
as a function of the $D$ momentum.}
\end{table}


\begin{references}
%
\bibitem{PDG}
C.\ Caso {\it et al.},
European Physical Journal {\bf C3}, 1 (1998),
and also the URL:\ http://pdg.lbl.gov.
%
\bibitem{CLEO}
CLEO Collaboration,
R.\ Godang {\it et al.},
Phys.\ Rev.\ Lett.\ {\bf 84}, 5038 (2000).
%
\bibitem{FOCUS}
FOCUS Collaboration,
J.\ M.\ Link {\it et al.},
Phys.\ Lett.\ B {\bf 485}, 62 (2000).
%
\bibitem{Ber00}
S.\ Bergmann {\it et al.},
Weizmann Institute report number WIS-8-00-DPP (2000),
hep-ph/0005181.
%
\bibitem{BLS}
See, for example,
G.\ C.\ Branco, L.\ Lavoura, and J.\ P.\ Silva,
{\it CP Violation}\, (Oxford University Press, Oxford, 1999).
%
\bibitem{early}
For early work see
K.\ M.\ Case,
Phys.\ Rev.\ D {\bf 103}, 1449 (1956);
M.\ L.\ Good,
Phys.\ Rev.\ D {\bf 106}, 591 (1957);
{\it ibid.\/} {\bf 110}, 550 (1958).
%
\bibitem{Fet96}
W.\ Fetscher {\it et al.},
Z.\ Phys.\ C {\bf 72}, 543 (1996).
%
\bibitem{Sch98}
Th.\ Schietinger,
{\it Regeneration of neutral $K$ mesons in carbon
at energies below 1GeV}\,
(Universit\"{a}t Basel, 1998),
PhD Thesis.
%
\bibitem{joao}
J.\ P.\ Silva,
SLAC preprint number SLAC-PUB-8499,
hep-ph/0007075.
%
\bibitem{definition}
Notice that the authors of reference \cite{Fet96} define
an $\Omega$ which is equal to half ours.
%
\bibitem{E791}
See, for example,
E791 Collaboration,
E.\ M.\ Aitala {\it et al.},
Phys.\ Rev.\ Lett.\ {\bf 77}, 2384 (1996).
%
\bibitem{fine-tunning}
There might be a caveat due to fine-tuning,
which is currently under investigation.
%
\bibitem{Gsp79diff}
A.\ Gsponer {\it et al.},
Phys.\ Rev.\ Lett.\ {\bf 42}, 13 (1979).
%
\bibitem{Roe77}
J.\ Roehrig {\it et al.},
Phys.\ Rev.\ Lett.\ {\bf 38}, 1116 (1977).
Here it is shown that experiments on carbon are consistent
with the ($\omega$) regge-pole-exchange
predictions.
%
\bibitem{Regge}
See, for example,
F.\ J.\ Gilman,
Phys.\ Rev.\ {\bf 171}, 1453 (1968);
V.\ Barger and D.\ Cline,
{\it Phenomenological Theories of High Energy Scattering}
(Benjamin, New York, 1968).
%
\bibitem{CPLEAR_disp}
CPLEAR Collaboration,
A.~Angelopoulos  {\it et al.}, 
Eur.\ Phys.\ J.\ C {\bf 10}, 19 (1999).
%
\bibitem{FOCUS-2}
This range is quoted by the FOCUS Collaboration in
J.\ M.\ Link {\it et al.},
hep-ex/000503.
%
\bibitem{Gsp79sum}
A.\ Gsponer {\it et al.},
Phys.\ Rev.\ Lett.\ {\bf 42}, 9 (1979).
%
\bibitem{Thomas}
The result quoted in reference \cite{Gsp79sum}
is
$[\sigma_T(K^0)+\sigma_T(\overline{K^0})]/2 \approx \sigma_T(K_L)$ 
= 23.5 mb ($A$/g mol$^{-1}$)$^{0.840}$,
where $\sigma_T$ are the total cross-sections.
Eq.~(\ref{eq:thomas}) is trivially obtained from this,
as shown to us by Th.\ Schietinger.
%
\bibitem{Dun95}
I.\ Dunietz,
Phys.\ Rev.\ D {\bf 52}, 3048 (1995).
%
\bibitem{Ben96}
See, for example,
M.\ Beneke, G.\ Buchalla, and I.\ Dunietz,
Phys.\ Rev.\ D {\bf 54}, 4419 (1996);
O.\ Schneider,
Institut de Physique des Hautes Energies report number
IPHE 2000-009,
hep-ex/0006006;
and references therein.
%
\end{references}
\end{document}